\documentclass[12pt]{article}
\usepackage{amsmath}
\usepackage{graphicx,psfrag,epsf}
\usepackage{enumerate}
\usepackage{natbib}
\usepackage{amssymb,amsmath,amsthm,bm} 
\usepackage{bbm}
\usepackage{hyperref}
\usepackage{natbib}
\setcitestyle{authoryear,open={(},close={)}}

\newcommand{\blind}{0}

\usepackage{color}
\usepackage{booktabs}
\usepackage{algorithm}
\usepackage[noend]{algpseudocode}
\usepackage{subcaption}
\usepackage{float}
\usepackage{caption}
\usepackage{algorithm,algcompatible,amsmath}
\usepackage{bbm}
\algnewcommand\INPUT{\item[\textbf{Input:}]}%
\algnewcommand\OUTPUT{\item[\textbf{Output:}]}%

\newcommand{\dd}{\mathrm{d}}

\renewcommand{\hat}{\widehat}
\renewcommand{\bar}{\overline}
\renewcommand{\tilde}{\widetilde}
\newtheorem{proposition}{Proposition}

\numberwithin{equation}{section}
\theoremstyle{plain}

\begin{document}

\def\spacingset#1{\renewcommand{\baselinestretch}%
{#1}\small\normalsize} \spacingset{1}

\if0\blind
{
  \title{\bf Bayesian Nonparametric Estimation for Point Processes with
  Spatial Homogeneity: A Spatial Analysis of NBA Shot Locations}
  \author{Fan Yin\\
    Microsoft \\
    Jieying Jiao \\
    Department of Statistics, University of Connecticut \\
    Guanyu Hu \\
    Department of Statistics, University of Missouri \\
    and \\
    Jun Yan \\
    Department of Statistics, University of Connecticut}
  \maketitle
} \fi


\if1\blind
{
  \bigskip
  \bigskip
  \bigskip
  \begin{center}
    {\LARGE\bf Title}
\end{center}
  \medskip
} \fi

\bigskip
\begin{abstract}
Basketball shot location data provide valuable summary information regarding
players to coaches, sports analysts, fans, statisticians, as well as players
themselves. Represented by spatial points, such data are naturally analyzed 
with spatial point process models. We present a novel nonparametric Bayesian
method for learning the underlying intensity surface built upon a
combination of Dirichlet process and Markov random field. Our method has the
advantage of effectively encouraging local spatial homogeneity when estimating 
a globally heterogeneous intensity surface. Posterior inferences are performed
with an efficient Markov chain Monte Carlo (MCMC) algorithm. Simulation studies
show that the inferences are accurate and that the method is superior compared
to the competing methods. Application to the shot location data of $20$ 
representative NBA players in the 2017-2018 regular season offers interesting
insights about the shooting patterns of these players. A comparison against the
competing method shows that the proposed method can effectively incorporate
spatial contiguity into the estimation of intensity surfaces.
\end{abstract}

\noindent%

\spacingset{1.45}

\section{Introduction}\label{sec:intro}
Quantitative analytics have been a key driving force for advancing modern
professional sports, and there is no exception for professional basketball
\citep{kubatko2007starting}. In professional basketball, analyses of
shooting patterns offer important insights about players' attacking styles
and shed light on the evolution of defensive tactics, which has aroused
substantial research interests from the statistical community
\citep[e.g.,][]{reich2006spatial, miller2014factorized, jiao2019bayesian,
  hu2020bayesiangroup}. As shot location data are naturally represented by
spatial points,  developments of novel methods for analyzing spatial point
patterns are of fundamental importance.

The literature on spatial point pattern data is voluminous \citep[see,
e.g.,][]{illian2008statistical, diggle2013statistical, guan2006composite,
  guan2010weighted, baddeley2017local, jiao2020heterogeneity}. The
most frequently adopted class of models in empirical research is
nonhomogeneous Poisson processes (NHPP), or more generally, Cox processes,
including log-Gaussian Cox process \citep{moller1998log}. Such parametric
models impose restrictions on the functional forms of underlying process
intensity, which can suffer from underfitting of data when there is a
misfit between the complexity of the model and the data available. In contrast,
nonparametric approaches provide more flexibility compared to parametric
modeling as the underfitting can be mitigated by using models with unbounded
complexity.

Several important features of the shot location data need to captured in any
realistic nonparametric method. First, near regions are highly likely to have
similar intensities. This makes that certain spatial contiguous constraints on
the intensity surface are desirable. Existing method mixture of finite mixtures
(MFM)
of nonhomogeneous Poisson processes \citep{geng2019bayesian} is lack of this
aspect. Second, spatial contiguous constraints should not dominate the intensity
surface globally \citep{hu2020bayesian, zhao2020bayesian}. This is because two
spatially disconnected regions that are sufficiently similar with respect to the
intensity values can still belong to the same cluster. For example, a player may
shoot equally frequently at the two corners due to the symmetry of the court,
which is not well accommodated by the penalized method \citep{li2019spatial}.
Finally, the extent to which the spatial contiguous affects the intensity
surface may differ from player to player, and needs to be learned from the data.

To address these challenges, we consider a spatially constrained Bayesian
nonparametric method for point processes to capture the spatial homogeneity
of intensity surfaces. Our contributions are three-fold. First, we develop a
novel nonparametric Bayesian method for intensity estimation of spatial point
processes. Compared to existing methods, the proposed approach is capable of
capturing both locally spatially contiguous clusters and globally discontinuous
clusters and the number of clusters. Second, an efficient Markov chain Monte
Carlo (MCMC) algorithm is designed for our model without complicated reversible
jump MCMC. Lastly, we gain important insights about the shooting behaviors of
NBA players based on an application to their shot location data.

The rest of the paper is organized as follows. In Section \ref{sec:data}, we
introduce and visualize the shot charts of several representative NBA players
from the 2017–2018 regular season. In Section~\ref{sec:model}, we introduce
spatial point processes for field goal attempts and develop a nonparametric
Bayesian method for capturing the spatial homogeneity of intensity
surface. Detailed Bayesian inference procedures, including a collapsed Gibbs
sampler and a post MCMC inference method, are presented in
Section~\ref{sec:bayesInf}.  Extensive simulation studies are reported in
Section~\ref{sec:simu}. The method is applied to $20$ key NBA players' shot
location data from the 2017-2018 regular season in
Section~\ref{sec:app}. Section~\ref{sec:disc} concludes with a discussion. For
ease of exposition, additional results are relegated to   
the Supplementary Material.

\section{NBA Shot Location Data}\label{sec:data}
Shot chart data for NBA players from the 2017--2018 regular season were
retrieved from the official website of NBA \url{stats.nba.com}. The data for
each player contain information about all his shots in regular season including
game date, opponent team name, game period when each shot was made (4~quarters
and a fifth period representing extra time),
minutes and seconds left, success indicator (0~represents missed and
1~represents made), action type (like ``Cutting dunk shot'', ``Jump
shot'', etc.), shot type (2-point or 3-point shot),
shot distance, and shot location coordinates. From the data, the half
court is positioned on a Cartesian coordinate system centered at the
center of rim, with $x$ ranging from $-250$ to 250 and $y$ ranging from
$-50$ to 420, both with unit of 0.1 foot (ft), as the size of an actual NBA
basketball half court is $50 \ \text{ft} \times 47 \ \text{ft}$.

\begin{figure}[th]
\centering
\includegraphics[width = \textwidth]{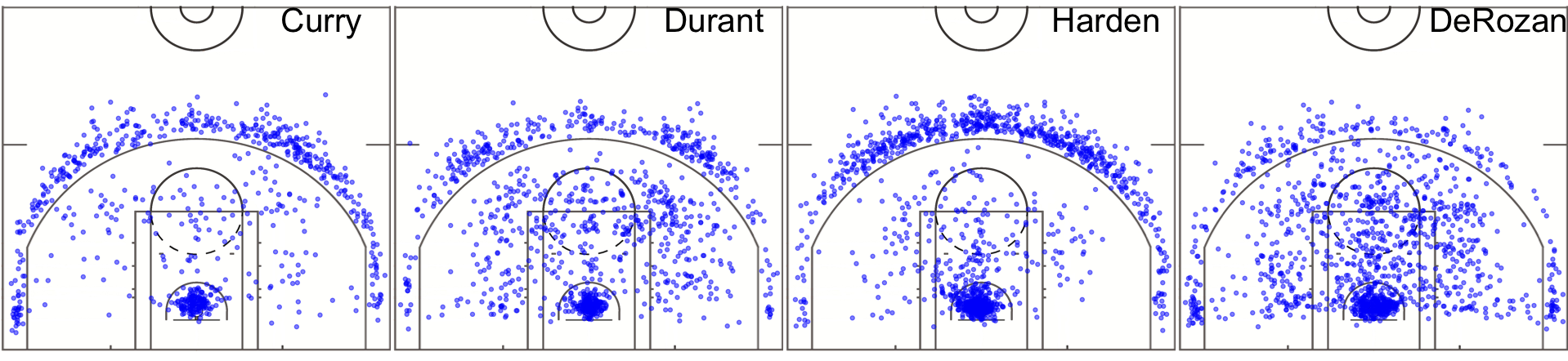}
\caption{Shot data Display. On half court image, each point represents
one shot. From left to right: Stephen Curry, Kevin Durant, James Harden, DeMar DeRozan.}
\label{fig:EDA}
\end{figure}

\begin{table}[tbp]
  \centering
  \caption{Shot data summary. Period is for the 1st, 2nd, 3rd, 4th quarter and the
    extra time.}
  \label{table:EADsummary}
  \begin{tabular}{cccc}
    \toprule
    Player & Shot Count & 2PT shot percentage ($\%$) & Period percentage ($\%$)\\
    \midrule
    
    Stephen Curry & 753 &  42.6 & (35.0, 20.6, 34.3, 9.7, 0.4)\\
    James Harden & 1306 &  50.2 & (28.7, 22.4, 27.9, 20.8, 0.3)\\
    Kevin Durant & 1040 &  66.5 & (30.8, 23.8, 30.6, 14.6, 0.3)\\
    DeMar DeRozan & 1274 & 79.9 & (29.1, 28.6, 33.3, 17.3, 1.6)\\
    \bottomrule
  \end{tabular}
\end{table}

We visualize and summarize the shot data of four key players, Stephen Curry,
James Harden, Kevin Durant and DeMar DeRozan.
Figure~\ref{fig:EDA} shows their field goal attempts' locations and
Table~\ref{table:EADsummary} summarizes their other
field goal attempts information. As we can see from the plots, most of the shots
are made either close to the rim or right outside the arc (i.e., 3-point line). This
is in line with the recent trend in the basketball development since it is more
efficient for players to pursue higher success rates near the rim or go after
higher rewards by making 3-point shots.

\section{Model}\label{sec:model}

\subsection{NHPP}
Spatial point process models provide a natural framework for capturing the
random behavior of event location data. Let
$\mathbf{S} = \{\bm{s}_1, \bm{s}_2, \dots, \bm{s}_N\}$ with
$\bm{s}_i = (x_i, y_i)$, $i=1,\ldots,N$,
be the set of observed locations in a pre-defined, bounded region
$\mathcal{B} \subseteq \mathcal{R}^{2}$. Let the underlying stochastic mechanism
that gives rise to the observed point pattern $\mathbf{S}$ be denoted as spatial
point process $\mathbf{Y}$. Process
$N_{\mathbf{Y}}(A) = \sum_{i=1}^{N} \mathbbm{1}(\bm{s}_{i} \in A)$
is a counting process associated with $\mathbf{Y}$, which counts the number of
points falling into area $A \subseteq \mathcal{B}$.

The NHPP model assumes conditionally independent event locations given the
process intensity $\lambda(\mathbf{s})$. For an NHPP, the number of events in
area~$A$, $N_{\mathbf{Y}}(A)$, follows Poisson distribution with rate parameter
$\lambda(A) = \int_{A} \lambda(\mathbf{s}) \dd \mathbf{s}$. In addition,
$N_{\mathbf{Y}}(A_{1})$ and $N_{\mathbf{Y}}(A_{2})$ are independent if two areas
$A_1 \subseteq \mathcal{B}$ and $A_2 \subseteq \mathcal{B}$ are disjoint. Given
the observed point pattern $\mathbf{S}$ on fixed region $\mathcal{B}$, the
likelihood of the NHPP model is
\begin{equation}
\label{eq:NHPP_lik}
\frac{\prod_{i=1}^{N} \lambda(\mathbf{s}_{i})}
{\exp(\int_{\mathcal{B}} \lambda(\mathbf{s}) d\mathbf{s})},
\end{equation}
where $\lambda(\mathbf{s}_{i})$ is the intensity function evaluated at location
$\mathbf{s}_{i}$. The NHPP reduces to a homogeneous Poisson process (HPP) when
$\lambda(\mathbf{s})$ is constant over the entire study region $\mathcal{B}$,
and it is synonymous with \emph{complete spatial randomness} (CSR).

\subsection{Nonparametric Bayesian Methods for NHPP}
As the CSR assumption over the entire study region rarely holds in real-world
problems, and to simplify the potentially overcomplicated problem induced by
complete non-homogeneity on intensity values, \citet{teng2017bayesian} proposed
to approximate the intensity function $\lambda(\bm{s})$ by a piecewise constant
function. Specifically, the study region $\mathcal{B}$ is partitioned into $n$
disjoint sub-regions and the intensity over each sub-region is assumed to be
constant. Let $A_{1}, A_{2}, \ldots, A_{n}$ be a partition of $\mathcal{B}$,
i.e., $\cup_{i=1}^{n} A_{i} = \mathcal{B}$ and $A_{i} \cap A_{j} = \emptyset,
\forall i \neq j$. For each region $A_{i}, i = 1, \ldots, n$, we have
$\lambda(\mathbf{s}) = \lambda_{i}, \forall \ \mathbf{s} \in A_{i}$. Therefore,
the likelihood~\eqref{eq:NHPP_lik} can be written as
\begin{equation}
\label{eq:NHPP_Poisson_lik}
\prod_{i=1}^{n} f_{\text{poisson}}(N_{\mathbf{Y}}(A_{i}) | \lambda_{i} \mu(A_{i})),
\end{equation}
where $\mu(A_{i} = \int_{A_{i}} 1 d\bm{s}$ is the area of sub-region $A_i$
$f_{\text{poisson}}(\cdot | \lambda)$ is the probability mass function of the
Poisson distribution with rate parameter $\lambda$. For ease of notation, we use
$N(A_{i})$ for $N_{\mathbf{Y}}(A_{i})$ in the sequel.

The heterogeneity in the intensity function across different sub-regions can be
naturally represented through a latent clustering structure. The conventional
finite mixture modeling framework \citep{mclachlan1988mixture,
  bouveyron2019model} assumes that the heterogeneity can be characterized by
a discrete set of subpopulations or clusters, such that the points located in
the sub-regions belonging to any given subpopulation tend to be produced by
similar intensities. The selection of the number of clusters (or components) in
finite mixture models are often recasted as statistical model selection problems
which can solved using information criteria \citep{fraley2002model} or
cross-validation \citep{fu2020estimating}, among others. Despite many successful
applications in empirical research, such model selection procedures are fraught
with in that they ignore the uncertainty in the number of clusters, which may in
turn lead to increased erroneous cluster assignments.

The Chinese restaurant process (CRP) \citep{pitman1995exchangeable,
  neal2000markov} mixture models is a class of Bayesian nonparametric approaches
that offer a powerful alternative to conventional finite mixture models. Under
CRP mixture models, the latent cluster indicator variables
$\mathbf{Z} = (Z_1, Z_2, \ldots, Z_n)$ are distributed according to a CRP, which
is defined through the following conditional distributions or a P\'{o}lya urn
scheme \citep{blackwell1973ferguson}
\begin{align}
\label{eq:polya_urn}
  \Pr(Z_i = c | Z_j, j< i; \alpha) \propto
  \begin{cases}
    |c|, & c \text{ is an existing cluster label} \ $c$, \\
    \alpha, & \text{otherwise},
  \end{cases}
\end{align}
where $|c|$ refers to the size of cluster labeled $c$, and $\alpha$ is the
concentration parameter of the underlying Dirichlet process (DP). Specifically,
Equation~\eqref{eq:polya_urn} implies that the trivial partition
$\left\{\left\{1 \right\}\right\}$ is obtained with probability~$1$ at the
beginning, and one new element is either added to one of the existing blocks of
the partition $\mathcal{C}_{n}$ with probability $|c|/(n+\alpha)$ or to the
partition $\mathcal{C}_{n}$ as a new singleton block with probability
$\alpha/(n+\alpha)$ in subsequent steps. More intuitively, we may think of
customers choosing tables in a restaurant, where the first customer always
chooses the first table and the $i$th customer chooses the first unoccupied
table with probability $\alpha/(n+\alpha)$ and an occupied table with
probability proportional to the number of customers currently sitting at that
table (i.e., how popular this table is), $|c|/(n+\alpha)$.

The CRP mixture models admit the uncertainty in the number of clusters and
circumvents the model selection problems by allowing the number of clusters and
cluster assignments to be inferred simultaneously. To facilitate statistical
inference, it is often helpful to consider the full conditional distributions of
$Z_{i}$, $i = 1, \ldots, n$ under the CRP,
\begin{align}
\label{eq:CRP_Z_full_cond}
  \Pr(Z_i = c | \bm{Z}_{-i}, \alpha) \propto
  \begin{cases}
    n_{c}(\bm{Z}_{-i}), & \text{at an existing cluster label} \ $c$, \\
    \alpha, & \text{at a new cluster},
  \end{cases}
\end{align}
where $\bm{Z}_{-i} = \{Z_j: j \ne i\}$, and
$n_{c}(\bm{Z}_{-i}) = \sum_{i=1,i \neq j}^{n} \mathbbm{1}(Z_{i}=c)$ is the
number of observations in cluster labeled~$c$ without the
$i$th~observation. That is, each $Z_{i}$ is either a new label with probability
proportional to $\alpha$, or an existing label~$c$ with probability proportional
to the observations assigned to cluster~$c$, $n_{c}(\bm{Z}_{-i})$.  The
concentration parameter~$\alpha$ controls the distribution of the number of
clusters, with smaller values of $\alpha$ favoring smaller number of clusters
\emph{a~priori}.

We formulate a Bayesian hierarchical CRP-NHPP model as
\begin{equation}
  \begin{split}
     Z_{1}, \ldots, Z_{n} &\sim \text{CRP}(\alpha),\\
     \lambda_1, \ldots, \lambda_K &\stackrel{i.i.d.}{\sim} 
     \mathrm{Gamma}(a, b)\\
     N(A_{i}) \mid  \bm{\lambda},  Z_{1}, \ldots, Z_{n} & \sim
     \text{Poisson}(\lambda_{Z_{i}}\mu(A_{i})), \ \text{independently for}
     \ i=1,\ldots,n,  
   \end{split}
  \label{eq:CRP_NHPP}
\end{equation}
where $K$ is the number of components of the piecewise constant function,
$\bm{\lambda} = (\lambda_1, \lambda_2, \dotsm \lambda_K)$ is the vector of
the unique values of the intensity function, $\text{Gamma}(a, b)$ is the gamma
distribution with shape $a$ and rate $b$, $(a, b, \alpha)$ are
hyperparameters for the prior distributions, and $i.i.d.$ stands for indepent
and identicall distributed. The hyperparameters can be
specified according to the \emph{a~priori} information available for the
practitioners.

An alternative model which will be used as a benchmark for comparison is the
MFM of NHPP (MFM-NHPP) \citep{geng2019bayesian}.
This model is built upon the MFM modeling
framework \citep{miller2018mixture} to mitigate the potential
\emph{inconsistency} in estimating the number of clusters caused by CRP
mixture. The MFM-NHPP model is
\begin{equation}
\begin{split}
  K &\sim p_{K},\\
  \lambda_{k} &\stackrel{i.i.d.}{\sim}
  \mathrm{Gamma}(a, b),\quad k = 1, 2, \dots, K,  \\
  (\pi_1,\ldots,\pi_k) \mid K = k & \sim
  \text{Dirichlet}_{k}(\alpha,\ldots,\alpha), \\
  Z_{1}, \ldots, Z_{n} \mid \pi_1, \ldots, \pi_k &\stackrel{i.i.d.}{\sim}
  \text{Categorical}_k(\pi_1, \ldots, \pi_K), \\
  N(A_{i}) \mid  \bm{\lambda},  Z_{1}, \ldots, Z_{n} & \sim
  \text{Poisson}(\lambda_{Z_{i}}\mu(A_{i})), \ \text{independently for}
  \ i=1,\ldots,n,  
 \end{split}
\label{eq:MFM_NHPP}
\end{equation}
where $p_{K}$ is the prior distribution on the number of clusters, that is, a
probability mass function on  $\mathbb{N} = \left\{1,2,\ldots \right\}$.
\citet{geng2019bayesian} adopt Poisson with mean~1 truncated to be positive for
$p_K$ as recommended by \citet{miller2018mixture}, which we follow here.

\subsection{Incorporating Spatial Homogeneity}

Spatial events typically obey the so-called first law of geography, ``everything
is related to everything else, but near things are more related than distant
things'' \citep{tobler1970computer}. This means spatial smoothness, also known
as spatial homogeneity. To incorporate such spatial homogeneity, we propose to
modify~\eqref{eq:CRP_Z_full_cond} by adding a multiplier that encourages the
customer to choose the table where many spatial neighbors are also sitting
at. In particular, we consider the following full conditionals
\begin{align}
\label{eq:SCCRP_Z_full_cond}
  & \Pr(Z_i = c | \bm{Z}_{-i}, \alpha, \eta, D) \\
  \propto & \begin{cases}
    n_{c}(\bm{Z}_{-i}) \exp \big(\eta \sum_{j \in \partial(i)}d_{ij}
    \mathbbm{1}(Z_j = c)\big),
    & c  \text{ is an existing cluster labeled} \ $c$, \\
    \alpha, & \text{otherwise},
  \end{cases}
\end{align}
where $D$ comprises the information on spatial distance and neighbor
relationships, $\partial (i)$ represents the set of spatial neighbors of
the~$i$th customer (observation), $d_{ij}$ denotes the spatial distance between
the~$i$th and the~$j$th customer (observation), and $\eta \geqslant 0$ is a
smoothing parameter controlling the relative weight of spatial homogeneity.

The full conditionals~\eqref{eq:SCCRP_Z_full_cond} can be specified by DP
mixtures (DPM) constrained by a Markov random field (MRF)
\citep{orbanz2008nonparametric}. Combining the NHPP with the MRF constrained
DPM, we have a MRF-DPM-NHPP model
\begin{equation}
  \begin{split}
     G  &\sim \text{DP}(\alpha,G_{0}) \\
     (\lambda_{1},\ldots,\lambda_{n}) &\sim
     M_{\eta, D}(\lambda_{1},\ldots,\lambda_{n}) \prod_{i=1}^{n} G(\lambda_{i}) \\
     N(A_{i}) \mid \lambda_1, \ldots, \lambda_n &\sim
     \text{Poisson}(\lambda_{i}) \ \text{independently for} \ i=1,\ldots,n,
  \end{split}
  \label{eq:MRF_DPM_NHPP}
\end{equation}
where $DP(\alpha, G_0)$ is a DP with base measure $G_0 \equiv \text{Gamma}(a,b)$
and concentration parameter $\alpha$, $G(\lambda_{i})$ is defined by a single draw from a DP \citep{ferguson1973bayesian}, and $M_{\eta,D}(\lambda_{1},\ldots,\lambda_{n})$ is a MRF with
full conditionals
\[
  M_{\eta,D}(\lambda_{i} | \bm{\lambda}_{-i}) = M_{\eta,D}(\lambda_{i} |
  \bm{\lambda}_{\partial (i)}) \propto \exp\big(\eta \sum_{j \in
    \partial(i)}d_{ij} \mathbbm{1}(\lambda_i = \lambda_j)\big)
\].
 
It is worth noting that the existence of joint distribution
$M_{\eta,D}(\lambda_{1},\ldots,\lambda_{n})$
is guaranteed by the Hammersley–-Clifford theorem
\citep{hammersley1971markov}. 


The definition of neighborhood $\partial (i)$ is
subject to the nature of the data and the modeler's choice.  Common choices
include the \emph{rook} contiguity (i.e., the regions which share a border of
some length with region $i$), and the \emph{queen} contiguity (i.e., the regions
which share a border of some length or even a point-length border with region
$i$). The smoothing parameter $\eta$ controls the extent of spatial homogeneity,
with larger values dictating larger extent of spatial homogeneity. The
MRF-DPM-NHPP model~\eqref{eq:MRF_DPM_NHPP} reduces to the CRP-NHPP
model~\eqref{eq:CRP_NHPP} when $\eta=0$.

\section{Bayesian Inference}\label{sec:bayesInf}
In this section, we present an efficient MCMC sampling algorithm for our
proposed method, post MCMC inference, and model selection criteria.

\subsection{A Collapsed Gibbs Sampler}
We introduce latent indicator variables
$\bm{Z} = (Z_1, \ldots, Z_n)$ and denote the parameters
in~\eqref{eq:MRF_DPM_NHPP} as
$\bm{\Theta} = \{\bm{\lambda}, \bm{Z}\}$.
The posterior density of $\bm{\Theta}$ is
\[
  \pi(\bm{\Theta} | \textbf{S}) \propto
  L(\bm{\Theta} | \mathbf{S}) \pi(\bm{\Theta})
\],
where $\pi(\bm{\Theta})$ is the prior density of $\bm{\Theta}$, and the
likelihood $L(\bm{\Theta} | \mathbf{S})$ takes the form of~\eqref{eq:NHPP_lik}.

We first derive the full
conditional distribution for each parameter as follows.
The full conditional probability that sub-region $A_i$
belongs to an existing component $c$, i.e.,
$\exists j \ne i,\, Z_j = c$, is
\begin{align}\label{eq:post_zexist_MRF_DPM_NHPP}
  \begin{split}
    \Pr(Z_i = c \mid
    \mathbf{S}, \bm{Z}_{-i}, \bm{\lambda})
    &\propto
    \frac{n_{c}(\bm{Z}_{-i})
      \exp \big(\eta \sum_{j \in \partial(i)}d_{ij} \mathbbm{1}(Z_j = c)\big)}
    {n-1+\alpha}
    \frac{(\lambda_{c} \mu(A_{i}))^{N(A_{i})}}
    {\exp(\lambda_{c}\mu(A_i))}.
  \end{split}
\end{align}
The full conditional probability that $A_i$ belongs to a new
component, i.e., $\forall j\ne i,\, Z_j \ne c$, is
\begin{align}\label{eq:post_znew_MRF_DPM_NHPP}
  \begin{split}
    &\phantom{ =\,\, } \Pr(Z_i = c\mid \mathbf{S}, \bm{Z}_{-i},
    \bm{\lambda}) \\
    & \propto
    \frac{\alpha}{n-1+\alpha}
    \int
    \frac{ (\lambda_{c} \mu(A_{i}) )^{ N(A_{i}))}}
    {\exp\left(\lambda_{c}\mu(A_i)\right)}
    \frac{b^a}{\Gamma(a)}\lambda_{c}^{a-1} e^{-b\lambda_{c}}
    \dd \lambda_{c}\\
    &=
    \frac{\alpha}{n-1+\alpha}
    \frac{b^a}{\Gamma(a)} \mu(A_{i})^{N(A_i)}
    \int
    \lambda_{c}^{N(A_{i})+a-1}e^{-(b+\mu(A_i))\lambda_{c}} \dd
    \lambda_{c}\\
    &= \frac{\alpha b^a \Gamma(N(A_{i})+a) \mu(A_{i})^{N(A_i)}}
    {(n-1+\alpha)
      (b+\mu(A_i))^{N(A_{i})+a} \Gamma(a)}.
  \end{split}
\end{align}
Combining~\eqref{eq:post_zexist_MRF_DPM_NHPP}
and~\eqref{eq:post_znew_MRF_DPM_NHPP}
gives the full conditional distribution of $Z_i$ in the following
Proposition.

\begin{proposition} \label{thm:z_MRF_DPM_NHPP}
Under the model and prior specification~\eqref{eq:MRF_DPM_NHPP}, the full
conditional distribution of $Z_i$, $i = 1, \ldots, n$, is
\begin{equation}
  \begin{split}
  &\phantom{ =\,\, }\Pr(Z_i = c\mid \mathbf{S}, \bm{Z}_{-i},\bm{\lambda}, \bm{\beta})\\
  &\propto
  \begin{cases}
    \displaystyle{
      \frac{n_{c}(\bm{Z}_{-i}) \exp \big(\eta \sum_{j \in \partial(i)}d_{ij} \mathbbm{1}(Z_j = c)\big) (\lambda_{c} \mu(A_{i}))^{N(A_{i})} }
      {\exp(\lambda_{c}\mu(A_i))}
    }
    & \exists j \ne i, \, Z_j = c \, \mbox{(existing label)},\\
    \displaystyle{
      \frac{\alpha b^a \Gamma(N(A_{i})+a) \mu(A_{i})^{N(A_i)}}
      {(b+\mu(A_i))^{N(A_{i})+a} \Gamma(a)}
    }
    & \forall j \ne i, \, Z_j \ne c \, \mbox{(new label)},
  \end{cases}
  \end{split}
  \label{eq:post_z_MRF_DPM_NHPP}
\end{equation}
where $\bm{Z}_{-i}$ is $\bm{Z}$ with $z_i$ removed, and
$\mu(A_i)$ is the area of region $A_i$.
\end{proposition}

For the full conditional distribution of
$\lambda_{k}$, only data points in the $k$th component should be
considered for simplicity.
The full conditional density of $\lambda_{k}$,
$k = 1, \ldots, K$, is
\begin{align}\label{eq:post_lambda_MRF_DPM_NHPP}
  \begin{split}
    q(\lambda_{k} \mid \mathbf{S}, \bm{Z}, \bm{\lambda}_{-k})
    &\propto
    \frac{\prod_{\ell:\bm{s}_\ell \in A_j,  Z_j = k}\lambda(\bm{s}_\ell)}
    {\exp(\int_{\bigcup_{j:Z_j = k}A_j}\lambda(\bm{s}) \dd \bm{s})}
    \lambda_{k}^{a-1}\exp\left(-b\lambda_{k}\right)\\
    &=
    \frac{\prod_{\ell:\bm{s}_\ell \in A_j, Z_j = k} \lambda_{k}}
    {\exp\left(\int_{\bigcup_{j:Z_j =  k}A_j}
      \lambda_{k} \dd \bm{s}\right)}
    \lambda_{k}^{a-1}\exp\left(-b\lambda_{k}\right)\\
    &\propto
    \lambda_{k}^{N_k+a-1}
    \exp\left(-\left(b + \sum_{j:Z_j = k}
        \mu(A_j)\right)\lambda_{k}\right),
  \end{split}
\end{align}
which is the kernel of
$\mbox{Gamma}\big(N_k+a, b+\sum_{j:Z_j = k}\mu(A_j) \big)$,
where $N_k = \sum_{\ell: \bm{s}_{\ell} \in A_{j}, Z_{j} = k} 1$ is the
number of data points in the sub-regions belonging to the $k$th component.

Algorithm~\ref{alg:MRF-DPM-NHPP} summarizes the steps of a Gibbs sampling
algorithm using the full conditional distributions
from~\eqref{eq:post_z_MRF_DPM_NHPP}--\eqref{eq:post_lambda_MRF_DPM_NHPP}.

\begin{algorithm}[tbp]
  \caption{Collapsed Gibbs sampler for MRF-DPM-NHPP.}
  \label{alg:MRF-DPM-NHPP}
  \begin{algorithmic}[1]
    \INPUT \hspace{0pt}\newline
    \indent Data: point locations $\bm{s}_{i}$, $i = 1, \ldots, N$;
    sub-regions and their neighbors $\{A_i, \partial (i):  i = 1, \ldots, n\}$. \newline
    \indent Prior hyperparameters : $a, b, \alpha$. \newline
    \indent Tuning parameter: $\eta$. \newline
    \indent Burn-in MCMC sample size: $B$, post-burn-in MCMC sample size: $L$. \newline
    \indent Initial values: $K$, $Z_1, \ldots, Z_n$,
    $\bm{\lambda} = (\lambda_1, \ldots, \lambda_{K})$, iter = 1.
    \newline
    \WHILE{iter $\leqslant B+L$}
    \STATE Update $\lambda_{k} | \cdot$, $k=1,\ldots,K$ as
      $$ \lambda_{k} | \cdot \sim \text{Gamma}\left(N_k+a, b+\sum_{j:Z_j =
    k}\mu(A_j) \right) $$
    \indent where $N_k = \sum_{\ell: \bm{s}_{\ell} \in A_{j}, Z_{j} = k} 1$ is the number of points belonging to the $k$th component.
    \STATE Update $Z_i | \cdot$, $i=1,\ldots,n$ following Proposition~\ref{thm:z_MRF_DPM_NHPP}.
    \STATE iter = iter + 1.
    \ENDWHILE
    \newline
    \OUTPUT Posterior samples $Z_{1}^{(l)},\ldots,Z_{n}^{(l)}$,
    $\bm{\lambda}^{(l)}$, $l=B+1,\ldots,B+L$.
  \end{algorithmic}
\end{algorithm}


Convergence check for the auxillary variables $(Z_1, \ldots, Z_n)$ can be done
with the help of the Rand Index (RI) \citep{rand1971objective}. The auxillary
variables themselves are nominal labels which cannot be compared from iteration
to iteration. The RI is the proportion of concordant pairs between two
clustering results with value of $1$ indicating the two results are exactly the
same. The trajectory of the RI for successive MCMC iterations provides a visual
check for convergence. Further, RI values closer to~1 indicate good agreement in
the clustering in the MCMC samples.

\subsection{Post MCMC Inference}\label{sec:post_mcmc}

We carry out posterior inference on the group memberships using Dahl's method
\citep{dahl2006model}, which proceeds as follows.

\begin{enumerate}
\item Define membership matrices
  $\mathcal{H}^{(l)} =(\mathcal{H}^{(l)}(i,j))_{i,j \in \left\{1,\ldots,n\right\} } =
  (\mathbbm{1}(Z_{i}^{(l)} = Z_{j}^{(l)}))_{n \times n}$,
  where $l = 1, \ldots, L$ indexes the number of retained MCMC draws after
  burn-in, and $\mathbbm{1}(\cdot)$ is the indicator function.
\item Calculate the average membership matrix
  $\bar{\mathcal{H}} = \frac{1}{L} \sum_{l=1}^{L} \mathcal{H}^{(l)}$
  where the summation is element-wise.
\item Identify the most \emph{representative} posterior sample as the one that
  is closest to $\bar{\mathcal{H}}$ with respect to the element-wise Euclidean
  distance
  $\sum_{i=1}^{n} \sum_{j=1}^{n} (\mathcal{H}^{(l)}(i,j) -
  \bar{\mathcal{H}}(i,j))^{2}$
  among the retained $l = 1,\ldots,L$ posterior samples.  
\end{enumerate}

Therefore, the posterior estimates of cluster memberships $Z_1,\ldots,Z_n$ and
model parameters $\bm{\Theta}$ can be based on the draws identified by Dahl's
method.

\subsection{Selection of Smoothing Parameter}\label{sec:eta_selection}
We recast the choice of smoothing parameter $\eta \geqslant 0$ as a model
selection problem. In particular, we consider the deviance information criterion
(DIC; \citet{spiegelhalter2002bayesian}), logarithm of the Pseudo-marginal
likelihood LPML; \citet{gelfand1994bayesian}) and Bayesian information
criterion (BIC; \citet{schwarz1978estimating}) as candidates.

The DIC for spatial point process can be derived from the standard DIC in a
straightforward manner as
\begin{equation*}
  \begin{split}
    \mbox{Dev}(\bm{\Theta}) &= -2  \left(\sum_{i = 1}^N
      \log\lambda(\bm{s}_i) - \int_{\mathcal{B}}\lambda(\bm{s})\dd
      \bm{s}\right),\\
    \mbox{DIC} &= 2\bar{\mbox{Dev}}(\bm{\Theta}) -
    \mbox{Dev}(\hat{\bm{\Theta}}),
  \end{split}
\end{equation*}
where $\bar{\mbox{Dev}}(\bm{\Theta})$ is the average deviance
evaluated using each posterior sample of $\Theta$, and
$\mbox{Dev}(\hat{\bm{\Theta}})$ is the deviance calculated using the
point estimation of parameter using Dahl's method.

The LPML for spatial point process can be approximated using the MCMC samples
\citep{hu2019new}
\begin{equation*}
  \begin{split}
    \widehat{\mbox{LPML}} &= \sum_{i = 1}^N
    \log\tilde{\lambda}(\bm{s}_i) - \int_{\mathcal{B}}
    \bar{\lambda}(\bm{s})\dd \bm{s},\\
    \tilde{\lambda}(\bm{s})_i &= \left(\frac{1}{M} \sum_{t = 1}^M
      \lambda(\bm{s}_i\mid \bm{\Theta}_t)^{-1}\right)^{-1},\\
    \bar{\lambda}(\bm{s}) &= \frac{1}{M}\sum_{t = 1}^M
    \lambda(\bm{s}\mid \bm{\Theta}_t),
  \end{split}
\end{equation*}
where $\bm{\Theta}_t$ denotes the $t$-th posterior sample of parameters
with a total length of $M$.

The BIC is derived naturally from its general definition
\begin{equation*}
  \begin{split}
    \mbox{BIC}(\bm{\Theta}) &= -2  \log L(\bm{\Theta}) + \hat{K} \log N,\\
    \log L(\bm{\Theta}) & = \sum_{i = 1}^N
      \log\lambda(\bm{s}_i) - \int_{\mathcal{B}}\lambda(\bm{s})\dd \bm{s},
  \end{split}
\end{equation*}
where $\hat{K}$ denotes the estimated number of components of the piecewise
constant intensity function.

\section{Simulation Studies}\label{sec:simu}

In this section, we report simulation studies to examine the performance of the
MRF-DPM-NHPP model and the proposed Gibbs sampling algorithm. In each setting,
we compare the results to that of MFM-NHPP To to show that the MRF-DPM-NHPP model
indeed leads to improvements.

\subsection{Design}\label{sec:simu_setup}
Consider a study region $\mathcal{B} = [0, 20] \times [0, 20]$ partitioned into
$n = 400$ squares of unit area, $\{A_i\}_{i = 1}^n$. The data generating model
was set to be $\mbox{NHPP}(\lambda(\bm{s}))$ with a piecewise constant intensity
$\lambda(\bm{s})$ over~$\mathcal{B}$. There settings were considered for
$\lambda(\bm{s})$; see Table~\ref{tb:simulation_settings}. The ``ground-truth''
intensity surfaces of the three settings are displayed in the leftmost column of Figure~\ref{fig:est_lambda_MRF_all}. The first two settings with the different numbers of clusters  are similar with the simulation setups in \citet{geng2019bayesian}. The third setting contains both spatially contiguous and discontinuous clusters. 
The point patterns were generated using the \texttt{rpoispp()} function from package \texttt{spatstat} \citep{baddeley2005spatstat}. For
each setting, we generated 100 replicates.

\begin{table}[tbp]
\centering
\caption{Simulation settings for the piecewise constant intensity function.}
\label{tb:simulation_settings}
\begin{tabular}{lccc}
\toprule
& $\bm{\lambda}$ & Number of components in $\bm{\lambda}$ & Number of grid boxes \\
\midrule
Setting 1 & $(0.2, 4, 12)$ & 3 & $(90, 211, 99)$ \\
Setting 2 & $(0.2, 1, 4, 8, 16)$ & 5 & $(80,80,80,80,80)$  \\
Setting 3 & $(0.2, 4, 10, 20)$ & 4 & $(90, 145, 66, 99)$\\
\bottomrule
\end{tabular}
\end{table}

The prior distributions were specified as in~\eqref{eq:MRF_DPM_NHPP}, with
hyperparameters $a = b = \alpha = 1$. The smoothing parameter
$\eta \geqslant 0$ took values on an equally-spaced grid
$\eta = \left\{0,0.5,\ldots,7.5,8 \right\}$,
of which the optimal value is chosen via the
model selection criteria introduced in Section~\ref{sec:eta_selection}.
The neighboring structure was defined based on rook contiguity, and we
treat all neighbors equally by letting $d_{ij} = 1, \forall j \in \partial i$.
Each MCMC chain was run for a total of
$4000$ iterations with random starting values, where the first $2000$ draws were
discarded as burn-in. The remaining $2000$ draws were thinned by~10 and
stored for posterior inference. We used Dahl's method \citep{dahl2006model} to
identify the most representative draw from the retained posterior draws as the
posterior point estimate.


\subsection{Results}\label{sec:simu_results}
We evaluate the results of simulation studies on the following aspects, (i)
probability of choosing the correct number of clusters, (ii) clustering accuracy
quantified by the RI, and (iii) estimation accuracy of the intensity surface.

Table~\ref{tb:Khat_meanRI_simulations} (left block) shows the proportion of times the true
number of components is identified under different
model selection criteria for each simulation setting. Obviously, $\eta=0$ never
recovered the true number of clusters, suggesting that taking spatial contiguity
information into account is crucial. For MRF-DPM-NHPP, BIC appears to be better
than DIC and LPML as the BIC-selected \emph{optimal} $\eta$ recovered the true
number of clusters considerably more frequently ($>70\%$). Although MFM-NHPP
seems to be very competitive in terms of identifying the true number of
components under setting~1, MRF-DPM-NHPP with smoothing parameter $\eta$
selected by BIC offers substantially better performance under setting~2 and
setting~3. A further investigation revealed that setting $\eta = 0$ always
produced overly large numbers of redundant clusters, while DIC and LPML failed
more gracefully with wrong numbers of clusters that often fall into the
approximate range (A histogram of~$\hat{K}$ is available in the Supplementary Material).

\begin{table}[tbp]
\centering
\caption{Proportion of times the true number of cluster is identified, and average RI across 100
  replicates for each simulation setting, under MFM-NHPP, and MRF-DPM-NHPP with
  $\eta=0$, optimal $\eta$ selected by BIC, DIC and LPML.}
\label{tb:Khat_meanRI_simulations}
\setlength\tabcolsep{5pt}
\begin{tabular}{l ccccc ccccc}
\toprule
& \multicolumn{5}{c}{Accuracy of $\hat{K}$} & \multicolumn{5}{c}{Average RI} \\
\cmidrule(lr){2-6}\cmidrule(lr){7-11}
& \multicolumn{4}{c}{MRF-DPM-NHPP} & \multicolumn{1}{c}{MFM-NHPP} & \multicolumn{4}{c}{MRF-DPM-NHPP} & \multicolumn{1}{c}{MFM-NHPP} \\
\cmidrule(lr){2-5}\cmidrule(lr){7-10}
& $\eta=0$ & BIC & DIC & LPML & & $\eta=0$ & BIC & DIC & LPML & \\
\midrule
Setting 1 & 0.00 & 0.75 & 0.21 & 0.22 & $\bm{0.95}$ & 0.619 & $\bm{0.974}$ & 0.890 & 0.891 & 0.905 \\ 
Setting 2 & 0.00 & $\bm{0.82}$ & 0.61 & 0.63 & 0.20 & 0.803 & $\bm{0.991}$ & 0.982 & 0.982 & 0.775 \\ 
Setting 3 & 0.00 & $\bm{0.71}$ & 0.10 & 0.11 & 0.56 & 0.739 & $\bm{0.992}$ & 0.939 & 0.942 & 0.870 \\ 
\bottomrule
\end{tabular}
\end{table}

To assess the clustering performance, we examine the average RI over the $100$
replicates. Because the ``ground-truth'' class labels are known in the
simulation studies, the RIs were calculated by comparing the MCMC iterations
with the truth as a measure of clustering accuracy.
As shown in Table~\ref{tb:Khat_meanRI_simulations} (right block),
MRF-DPM-NHPP with smoothing parameter $\eta$ selected by BIC yields the highest
clustering accuracy. Although Despite being more capable of identifying the true number
of clusters, the clustering accuracy of MFM-NHPP is worse than that of
MRF-DPM-NHPP with BIC under setting 1, which suggests that MFM-NHPP might happen
to get the number of clusters right by allocating the regions into wrong
clusters. For the remainder of this paper, we focus on the results that
correspond to optimal $\eta$ selected by BIC.

\begin{figure}[tbp]
\centering
\includegraphics[trim={4.5cm 0cm 4.2cm 0cm}, clip=true, width=\textwidth]{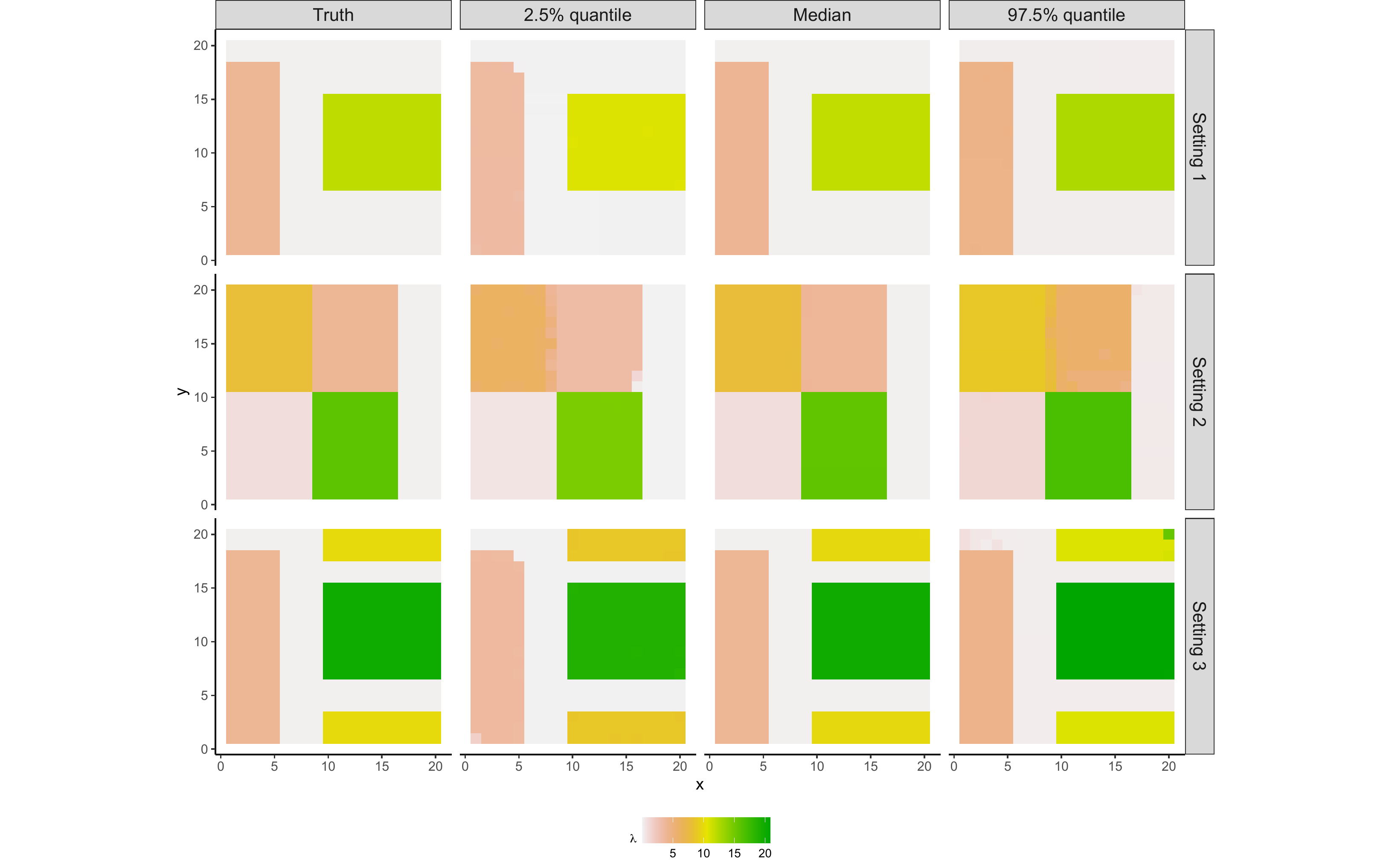}
\caption{Simulation configurations for intensity surfaces, with fitted intensity surfaces. Element-wise median and quantiles are calculated out of 100 replicates.}
\label{fig:est_lambda_MRF_all}
\end{figure}

\begin{figure}[tbp]
\centering
\includegraphics[trim={3.5cm 0cm 3.2cm 0cm}, clip=true, width=\textwidth]{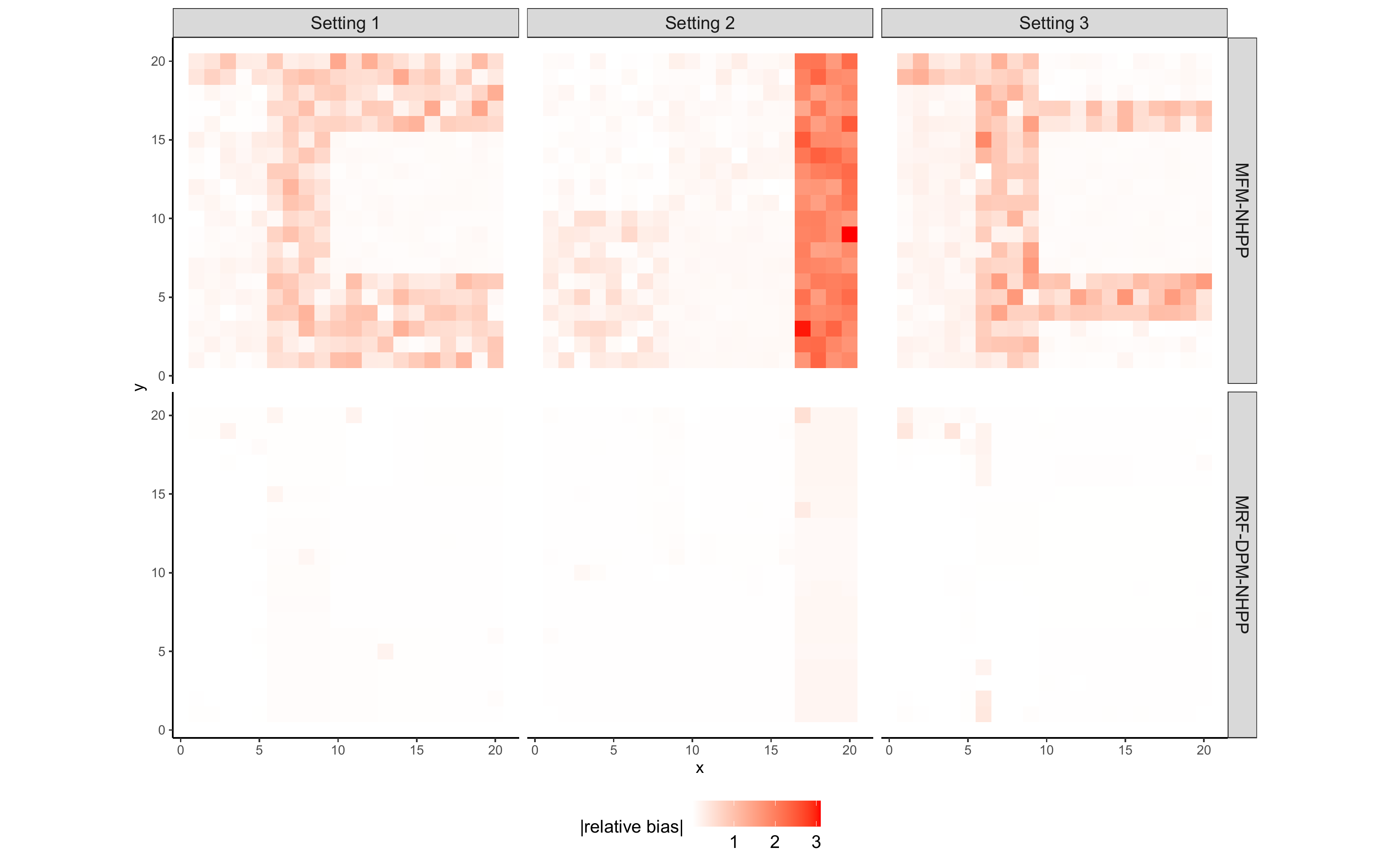}
\caption{Absolute value of relative bias of element-wise posterior mean estimates for intensity surfaces.}
\label{fig:est_lambda_relative_bias}
\end{figure}


We next summarize accuracy in estimating the intensity
surfaces. Figure~\ref{fig:est_lambda_MRF_all} displays the averages of
the median, 2.5th percentile, and 97.5th percentile of the estimated
intensity surface obtained with the optimal~$\eta$ selected by BIC from the 100
replicates, in comparison with the true surfaces, for the three settings. The median
surface agrees with true surface well in all three settings. The average
surfaces of the 2.5th and 97.5th percentiles of the 100 replicates have higher
uncertainties occasionally at the boundaries where the true intensities jump,
but in general are not
far from the true surfaces. Figure~\ref{fig:est_lambda_relative_bias} shows the
absolute value of relative bias of element-wise posterior mean estimates under
the MFM-DPM-NHPP model and the MFM-NHPP model.
The proposed method leads to substantially smaller bias than the competing method,
especially for sub-regions with low true underlying intensity values
and/or sub-regions at the boundaries. The advantage of the proposed method comes
from leveraging the spatial contiguity in the presence of spatial homogeneity.

The overlaid traceplots of RI (available in Supplementary Material) indicate that the chains converge very fast, and
stablize after $2000$ iterations, regardless of the settings. This observation
justifies our choice of the MCMC setting.

In summary, the simulation studies confirm the advantages of the
MRF-DPM-NHPP model and the validity of the proposed Gibbs sampling
algorithm. The results also suggest that BIC is better than DIC and LPML in
selecting the smoothing parameter $\eta$ in the studied settings.
Compared to the benchmark MFM-NHPP model, the MRF-DPM-NHPP
model is superior in clustering accuracy and has an advantage on identifying the
true number of components under more complex settings.

\section{Professional Basketball Data Analysis}\label{sec:app}

We applied the MRF-DPM-NHPP model to study the shot data for NBA players in the
2017-2018 NBA regular season described in Section~\ref{sec:data}. In particular,
we focus on $20$ all-star level players that are representative of their
positions (Table ~\ref{tb:player_info}).
The study region is a rectangle covering the first 75\% of the half court
($50 \ \text{ft} \times 35 \ \text{ft}$) as the shots made outside this region
are often not part of the regular tactics. This rectangle was divided into
$50 \times 35 = 1750$ equally-sized grid boxes of
$1\text{ft} \times 1\text{ft}$. For each player, we run an MCMC
with $\eta \in \{0,0.5,\ldots,6.5,7\}$ for $4000$ iterations, where
the first $2000$ were discarded as burn-in and the remainder was thinned by~10.

\begin{table}[tbp]
\centering
\caption{Basic information (name and the preferred position) of players and the
  number of clusters given by MRF-DPM-NHPP with the smoothing parameter selected
  by BIC, and by MFM-NHPP. Player positions: point guard (PG), shooting
  guard (SG), small forward (SF), power forward (PF), center (C).}
\label{tb:player_info}
\begin{tabular}{c c rr  r}
   \toprule
\multicolumn{1}{c}{} & \multicolumn{1}{c}{} & \multicolumn{2}{r}{MRF-DPM-NHPP} & \multicolumn{1}{r}{MFM-NHPP} \\
\cmidrule(lr){3-4}
 Player & Position & $\hat{K}_{\text{BIC}}$ & $\hat{\eta}_{\text{BIC}}$ & $\hat{K}$ \\ 
\midrule
  Joel Embiid & C & 12 & 2.5 & 9 \\ 
  Dwight Howard & C & 6 & 2.5 & 11 \\ 
  DeAndre Jordan & C & 6 & 4.0 & 4 \\
  Karl-Anthony Towns & C & 8 & 2.5 & 9 \\ 
  \hline
  LaMarcus Aldridge & PF & 7 & 2.5 & 17 \\ 
  Giannis Antetokounmpo & PF & 6 & 3.0 & 10 \\ 
  Blake Griffin & PF & 8 & 2.5 & 10 \\ 
  Kristaps Porziņģis & PF & 5 & 2.5 & 9 \\ 
  \hline
  Stephen Curry & PG & 5 & 3.0 & 3 \\
  Kyrie Irving & PG & 5 & 3.0 & 9 \\ 
  Damian Lillard & PG & 6 & 3.0 & 6 \\ 
  Chris Paul & PG & 8 & 2.5 & 5 \\ 
  \hline
  Jimmy Butler & SF & 5 & 3.0 & 9 \\ 
  Kevin Durant & SF & 9 & 3.0 & 13 \\
  Paul George & SF & 6 & 3.5 & 8 \\ 
  LeBron James & SF & 6 & 3.0 & 8 \\ 
  \hline
  DeMar DeRozan & SG & 8 & 3.0 & 10 \\ 
  James Harden & SG & 8 & 3.0 & 11 \\ 
  Klay Thompson & SG & 6 & 5.0 & 11 \\ 
  Russell Westbrook & SG & 5 & 3.5 & 10 \\ 
   \bottomrule
\end{tabular}
\end{table}

Table~\ref{tb:player_info} summarizes the optimal $\eta$ selected by BIC and the
resulting number of clusters. None of the selected $\hat{\eta}$
lies on the boundary, which assures the validity of candidate values of
$\eta$. For comparison, the number of clusters from the MFM-NHPP model under the same MCMC setting is also included, and we note that MFM-NHPP leads to higher numbers of clusters for most of the players than that of MFM-DPM-NHPP. 

\begin{figure}[htp]
    \centering
    \includegraphics[width=\textwidth]{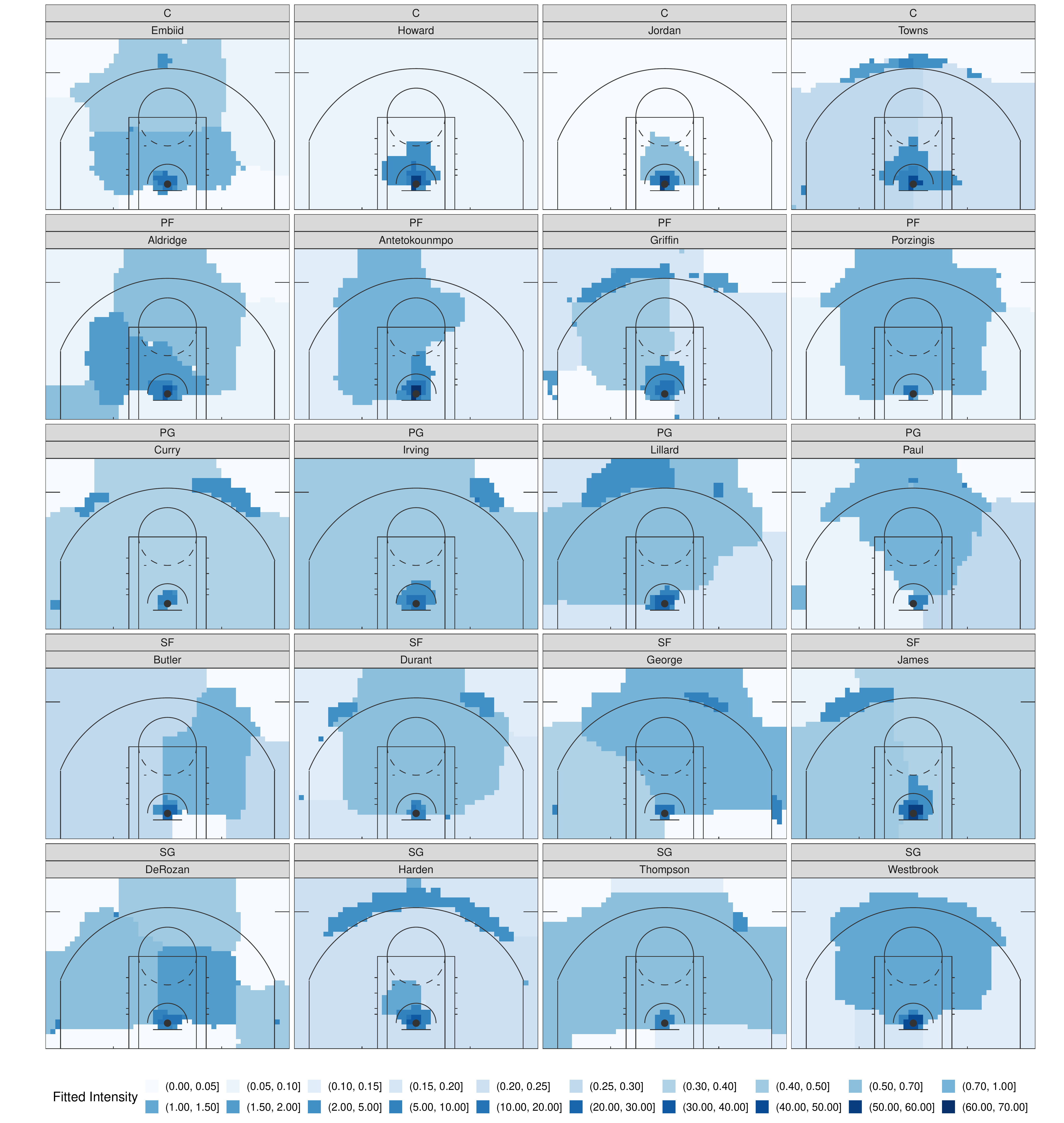}
    \caption{Estimated shooting intensity surfaces of selected players based on MRF-DPM-NHPP.}
    \label{fig:real_data_MRF_results}
\end{figure}

\begin{figure}[htp]
    \centering
    \includegraphics[width=\textwidth]{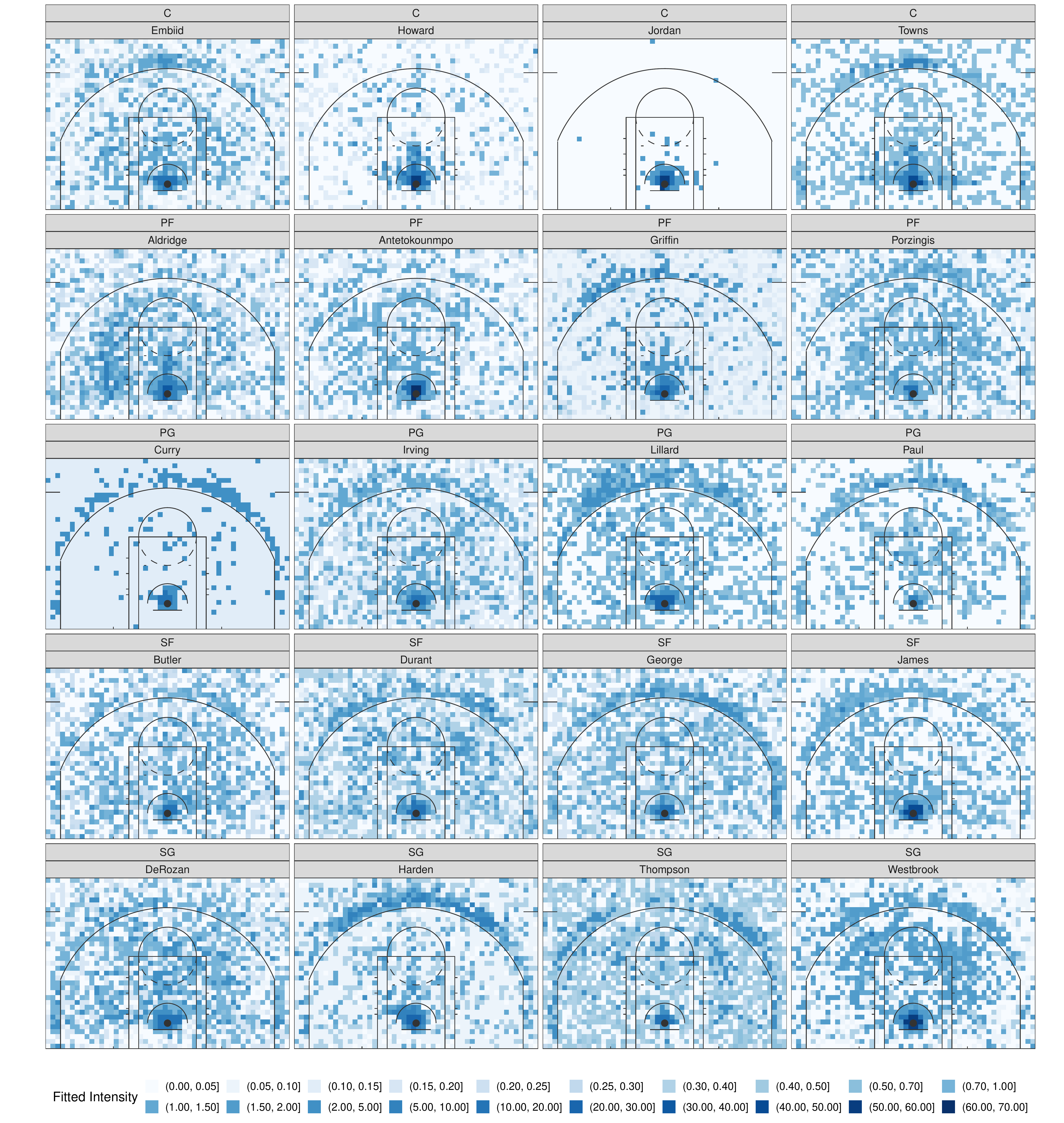}
    \caption{Estimated shooting intensity surfaces of selected players based on MFM-NHPP.}
    \label{fig:real_data_MFM_results}
\end{figure}


Figure~\ref{fig:real_data_MRF_results} and
Figure~\ref{fig:real_data_MFM_results} show the estimated shooting intensity
surfaces of selected players under MRF-DPM-NHPP and MFM-NHPP,
respectively. Compared to the results of MFM-NHPP, it is clear that the
MRF-DPM-NHPP model is capable of
capturing distant regions that share similar shooting intensities while
preserving the spatial contiguity, which greatly facilitates the interpretability. 
Taking Paul George as an example, the estimated shooting intensity surface
yielded by MFM-NHPP appears to be too scattered to highlight his preferred
shooting regions; the results from the MRF-DPM-NHPP model, however, shows much
clearer pattern.

More interesting observations are seen from the estimated shooting intensity
surfaces, and we summarize these observations by the preferred positions of
selected players. Among those players with preferred position as center, DeAndre
Jordan and Dwight Howard rarely make shots outside the low post, while the
latter seems to have made more shots from the regions between short corner and
the restricted area. On the contrary, Joel Embiid and Karl-Anthony Towns are
more versatile as attackers in terms of their shot locations --- Joel Embiid can
attack from low post, high post, top of the key as well as the \emph{point}
(i.e., right outside the middle of the arc); Karl-Anthony Towns' shots are mainly
initiated either from the low block or outside the arc (right corner and from
point to the wing).

The selected power-forward (PF) players show fairly different shooting
styles. The shot locations of Kristaps Porziņģis are very similar to those of
Joel Embiid, and Kristaps Porziņģis seems to be less confined to shooting from
low post regions compared to Joel Embiid. Both Giannis Antetokounmpo and
LaMarcus Aldridge all make substantial amounts of mid-range shots and seldomly
make three-point shots, but it is worth highlighting their differences as
Giannis Antetokounmpo seems to be more inclined to make shots from the right
while LaMarcus Aldridge's mid-range shots are more spread. Interestingly, the
former champion of slam dunk contest, Blake Griffin has higher intensity of
shooting outside the arc (in particular, from the right corner, and the regions
between the wing and the point).

The selected small-forward (SF) players show versatile shot locations but they
differ substantially in their three-point shot locations and the intensity of
making shots around restricted area. Speaking about the three-point shots, Kevin
Durant prefers shooting around left and right wings, both Paul George and Jimmy Butler prefer
shooting around the right corner but the former is clearly more comfortable with launching long-range shots, while LeBron James prefers
shooting around the left wing. Compared to the other two SF players, LeBron
James have higher intensity of making shots around the restricted area.

The difference in the shooting patterns among backcourt (PG and SG) players is
substantial. James Harden, Stephen Curry, Damian Lillard and Kyrie Irving all
launch considerable amounts of shots within the restricted area and outside the
arc, while James Harden makes shots in almost all regions from right wing to
left wing right outside the arc, Stephen Curry and Kyrie Irving make more shots
around left wing rather than right wing, Damian Lillard makes more shots around
right wing rather than left wing. Compared to the former three players,
Chris Paul, Russell Westbrook, DeMar DeRozan and Klay Thompson make more mid-range shots, but
from different angles. Specifically, Russell Westbrook makes shots almost
everywhere in the middle, Chris Paul's shots are mainly located in a
sector-shaped area in the middle, Demar DeRozan's shots are more spread to
the corners, while Klay Thompson's shots are almost evenly distributed across the entire study region.

Admittedly, the above analysis is far from being exhaustive. We believe,
however, that basketball professionals may leverage the proposed
method to better understand the shooting patterns of the players and, therefore,
design highly targeted offense and defense tactics.

\section{Discussion}\label{sec:disc}
The NBA shot location data appear to be modeled by the
spatially constrained nonparametric Bayesian model, MRF-DPM-NHPP, reasonably
well incorporating local spatial homogeneity. Building upon a combination of
Dirichlet process and Markov random field, the proposed method relies on a smoothing
parameter $\eta$ to effectively control the relative contribution of local
spatial homogeneity in estimating the globally heterogeneous intensity
surface.  Statistical inferences are facilitated by a Gibbs sampling
algorithm.  Selection of the smoothing
parameter $\eta$ is casted as a model selection problem which is handled using
standard model selection criteria. Simulation studies show the accuracy of the
proposed algorithm and the competitiveness of the model relative to the
benchmark MFM-NHPP model \citep{geng2019bayesian} 
under several settings in which spatial contiguity is present in
the intensity surface. In application to the shot locations of NBA players, the
model effectively captures spatial contiguity in shooting intensity surfaces,
and provide important insights on their shooting patterns which cannot be
obtained from the MFM-NHPP model.

There are several possible directions for further investigation.
More sophisticated definition of neighborhood (e.g.,
higher-order neighborhood, incorporating covariates) than the rook contiguity,
which was used in this study and found to be sufficient here, may be
useful for more complex data structure. BIC was found to perform well for the
purpose of selecting smoothing parameter $\eta$, but it is of substantial
interest to develop a fully automated procedure that enables the smoothing
parameter to be inferred along with the intensity values and the group
membership indicators through a single MCMC run. The NBA players
shot pattern modeling admits a natural partition for the region of interest.
In general settings, however, it is worth investigating how to effectively
partition the space such that the piecewise constant assumption is more
plausible. As the number of parameters is proportional to the number of grid
boxes, developments of more scalable inference algorithms (e.g., variational
inference) are critical for finer grid. Finally, building a group learning
model with pooled data from multiple players merits future research from both
methodological and applied perspectives.

\section*{Acknowledgements}
The authors would like to thank Dr.~Yishu Xue for sharing the R code of data
visualization.

\bibliographystyle{abbrvnat}
\bibliography{main}

\end{document}